\documentclass[runningheads]{llncs}
\usepackage{graphicx}
\usepackage{hhline}

\usepackage[hyphens]{url}
\usepackage{hyperref}
\usepackage[hyphenbreaks]{breakurl}

\begin{document}
\title{Solving Interactive Fiction Games via Partial Evaluation and Bounded Model Checking}
\titlerunning{Solving IF Games via Partial Evaluation and Bounded Model Checking}
\author{Martin Mariusz Lester\orcidID{0000-0002-2323-1771}}
\authorrunning{M M Lester}
\institute{Department of Computer Science, University of Reading, United Kingdom
\email{m.lester@reading.ac.uk}
}
\maketitle              % typeset the header of the contribution
\begin{abstract}
%The abstract should briefly summarize the contents of the paper in
We present a case study on using program verification
tools, specifically model-checkers for C programs, to solve simple interactive fiction games from around 1980.
Off-the-shelf model-checking tools are unable to handle the games in their original form.
In order to work around this, we apply a series of program transformations
that do not change the behaviour of the program.
An interesting aspect of these games is that they
use a simple, interpreted language to script in-game events.
This turns out to be the most difficult part of the program for
verification tools to handle;
we tackle this using partial evaluation.
Our case study thus provides some insights that
are applicable more generally to verification and analysis of programs
that interpret scripting languages.

To the best of our knowledge, this is the first example of
a commercially released game being solved by application of
a program model-checker to the game's code.

\end{abstract}
\section{Introduction}
Interactive fiction games, also known as text adventures, are an early form of computer game.
A game takes the form of a dialogue between the computer and a human player.
The player controls a character in a virtual world, often inspired by science-fiction or fantasy literature.
The computer describes the player's location.
The player gives textual commands to the computer, such as ``GET KEY'', ``GO NORTH'' or ``OPEN DOOR''.
The computer executes these commands in the game world and describes their outcome.
The sequence repeats until the player wins the game (for example, by finding some treasure)
or loses the game (often by dying).
We present a case study on using automated program verification tools,
in particular CBMC, a bounded model-checker for C programs, to solve these games.

In outline, we insert an assertion into the game that is violated on completion of the game,
then ask a model-checker to prove that the violated assertion is unreachable.
If the game can be completed, the model-checker ought to fail,
at which point the counterexample it produces shows how to complete the game.
However, despite the age and apparent simplicity of the games we considered,
we find that leading program model-checkers are unable to solve even very simple games.

To address this, we apply a sequence of program transformations,
which do not change the behaviour of the game programs,
but enable a model-checker to handle them.
Most importantly, we use a form of partial evaluation to tackle path
explosion in the interpreter loop in the game's scripting engine.
Furthermore, so that we can take advantage of program transformations already implemented
as program optimisations, we use the LLVM/Clang compiler.
However, as LLVM generates intermediate code, but we wish to use tools that take C programs as input,
we use the LLVM C Backend to generate a C program from the optimised intermediate code.

We explain our motivation for this case study, including why we believe this is an interesting problem, in
Section~\ref{sec:motivation}.
Because the game engine we consider uses an embedded, interpreted scripting language,
we argue that the techniques are applicable to other programs with scripted behaviour.

In Section~\ref{sec:background}, we give some background information on the problem,
including some details about the class of games we consider,
namely Scott Adams Grand Adventures~\cite{saga} (SAGA).
Of particular relevance is that the games have a large but finite state space.

Next, in Section~\ref{sec:adapt}, we explain how we posed solution of a game as a verification problem
and, in Section~\ref{sec:pe}, how we used partial evaluation to enable its solution using automated verification tools.
We focus on application of the established bounded model-checker CBMC~\cite{DBLP:conf/tacas/ClarkeKL04},
but we also considered tools that ranked highly in the recent SV-COMP 2020
Competition on Software Verification~\cite{DBLP:conf/tacas/Beyer20}.
As a result, we were able to solve a small tutorial game
and the Adventureland Sampler using CBMC.

In Section~\ref{sec:res} we give some quantitative results from our case study
and in Section~\ref{sec:eval} we reflect on the lessons learnt.

We consider related work, both on the problem of applying verification tools
to scripting languages, and on use of automated tools and techniques on interactive fiction games,
in Section~\ref{sec:related}.
This includes our custom tool ScAmPER, which tackles the same class of games
using a very fast but stupid heuristic search.

Finally, in Section~\ref{sec:conc}, we conclude with a summary of what we achieved.
The artifact accompanying this paper includes our tool ScAmPI (Scott Adams
Partially-evaluated Interpreter),
based around our modified version of the game engine,
which allows the technique to be applied to any SAGA game.

The contributions of this paper are as follows:
\begin{enumerate}
\item We have developed the first tool that automatically translates the problem of
solving a SAGA game accurately into a model-checking problem.
\item We show how to apply partial evaluation to the interpreter
loop for the game engine's scripting language,
which makes the verification problem solvable using currently available C
program model-checkers.
Our techniques should be applicable to interpreters for other scripting languages.
\item To our knowledge, this is the first successful example of a verification tool
being applied to a game engine for a commercially released computer game in order to solve the game.
\item More generally, our case study serves as a detailed account of
how to solve the problems encountered when using a bounded model-checker on a realistic
C program that, while still relatively small, is larger than the toy examples found in many research papers.
\end{enumerate}

\section{Motivation}
\label{sec:motivation}

The successful application of machine learning to playing old Atari video games~\cite{DBLP:journals/corr/MnihKSGAWR13,mnih2015human}
has captured the interest of both researchers and the general public.
They are an appealing application of research for a number of reasons:
\begin{itemize}
\item the public understand what a computer/video game is and what its objective is;
\item the player's score is an easily available metric for assessing success automatically;
\item older games are simple enough that machine learning techniques can be very successful;
\item older games, while still under copyright, are easy to obtain for use in research
and provide a wide range of realistic examples for evaluating a new technique.
\end{itemize}

However, there has been relatively little research of a comparable nature that uses techniques from formal verification.
With access to the internal workings of a game, verification techniques ought (in theory) to be able to produce better results,
although plays produced by verification tools might seem strange to a human observer,
as they could use information about the game state that is invisible to a human player.
One of the main problems in verification is state space explosion,
but the limited memory of older computer systems mitigates this,
meaning that verification techniques are more likely to be effective in this domain than with modern applications software.

Solutions to the problem of using verification techniques to play video games have broader applications
for two main reasons.
Firstly, solution of games by humans typically involves a mixture of high-level and low-level reasoning.
For example, navigating through a maze of rooms involves high-level reasoning about the connections between rooms
and low-level reasoning about how to move from one end of a room to another.
An effective verification technique for this kind of problem will need either to combine these kinds of reasoning,
or to be sufficiently good at low-level reasoning that high-level reasoning is not needed.
The same is true of many problems in program analysis.

Secondly, source code for older games is not usually available, so they must be analysed at the binary level.
The binary code may be for a processor for which there are no existing tools,
so either a new tool must be made, requiring great expenditure of effort,
or an existing tool can be applied to an emulator for the processor,
which introduces a further layer of complexity to the verification process.
Even then, the game might have been written in BASIC or another interpreted language,
resulting in multiple layers of emulation or interpretation to analyse.
Yet the same problems arise when trying to analyse a program written in an unusual language
or when trying to analyse compiled programs for security purposes;
consider, for example, analysis of a program written in OCaml but compiled to JavaScript.

\section{Choice of Case Study}
\label{sec:background}

We chose to tackle interactive fiction games for two reasons.
Firstly, measured by number of player actions,
solutions to interactive fiction games are often short.
A game might be solvable with under a hundred commands;
an arcade game might need thousands of joystick inputs.
This limits the depth of solutions that must be considered,
although the range of possible commands means the branching factor is large.
Secondly, many interactive fiction games published by the same company
consisted of a generic interpreter and a game-specific database.
The database describes the contents of the game world and
contains script code that is executed when certain events are triggered.
Many interpreters have now been rewritten in C,
allowing the games to be played on modern computers.
This meant that we only had to deal with one level of emulation or interpretation
(for the in-game scripting language)
and could use off-the-shelf C verification tools.

The most popular of the generic interactive fiction interpreters was Infocom's Z-Machine.
%Many of the iconic games of the 1980s were written for this virtual machine
%and it remains the most popular format for distribution of interactive fiction games today.
However, its flexibility, which allows games to include their own customised command parsers,
makes it a difficult target for verification.
Instead, we decided to tackle an earlier and simpler format,
namely SAGA games and the corresponding open-source interpreter ScottFree~\cite{scottfree}.
This format supports only very limited scripting, which makes the behaviour of the games far less dynamic.
%Games consist of a fixed number of rooms and objects.
%Commands consist of one or two words (a verb and a noun), which are taken from a fixed list.
%The player can move between rooms and pick up and drop objects.
%A limited scripting system allows the player to trigger special events by entering certain rooms or typing certain commands, provided certain conditions are satisfied.
%Examples of conditions that can be checked include the location of a certain object or the value of a finite number of flags and bounded counters.
%Examples of events include moving objects, moving the player and displaying messages.
Pseudocode for a simplified version of the game engine and some examples of scripted events from
the game Tutorial 4 (described below) are shown in Fig.~\ref{fig:engine}
(reproduced from Appendix A of~\cite{DBLP:conf/tap/Lester20}).

\begin{figure}[p]
\scriptsize
\begin{verbatim}
// Game engine
while (not game_over) {
    print(current_room.description());
    execute_automatic_scripts();
    (verb, noun) = parse_player_input();
    if (scripted_action(verb, noun)) {
        execute_scripted_action(verb, noun);
    }
    else if (verb == "go") {
        current_room = current_room.exit[noun];
    }
    else if (verb == "get" and items[noun].location == current_room) {
        items[noun].location = carried;
    }
    else if (verb == "drop" and items[noun].location == carried) {
        items[noun].location = current_room;
    }    
}

// Example scripted actions
// Action 0:
if (verb == "score") {
    print_score();
}
// Action 1:
else if (verb == "inventory") {
    print_inventory();
}
// Action 2:
else if (verb == "open" and noun == "door" and items["locked door"].location == current_room and
    items["key"].location != carried and items["key"].location != current_room) {
    print("It's locked.");
}
// Action 3:
else if (verb == "open" and noun == "door" and items["locked door"].location == current_room) {
    swap(items["locked door"].location, items["open door"].location);
}
// Action 4:
else if (verb == "go" and noun == "door" and items["open door"].location == current_room) {
    current_room = rooms["cell"];
}

// Example automatic scripts
// Action 5:
if (items["vampire"].location == current_room and items["cross"].location != carried) {
    print("Vampire bites me! I'm dead!);
    exit();
}
// Action 6:
if (items["vampire"].location == current_room and items["cross"].location == carried) {
    print("Vampire cowers away from the cross!);
}


\end{verbatim}
\caption{Pseudocode for the structure of the game engine and some scripted events.
Scripted events are taken from Tutorial 4 of Mike Taylor's Scott Adams Compiler~\cite{sac}.
The functions invoked by actions 0 and 1, which display the player's score and inventory
(list of items carried), are built into the game engine.}
\label{fig:engine}
\end{figure}

Our case study focuses on three specific games.
\emph{Tutorial 4} was taken from Mike Taylor's Scott Adams Compiler,
a relatively recent (2006) tool for creating and editing SAGA games.
This is the smallest non-trivial game we could find; it includes 2
scripted puzzles and the possibility of winning
by finding a treasure or losing by being bitten by a vampire.
\emph{Adventureland} (1978) was the first text adventure released for a
microcomputer. The goal is to collect 13 treasures. The game was highly
commercially successful and released on 13 different platforms.
\emph{Adventureland Sampler} (1979) was a cut-down version of Adventureland,
sold at a lower price. The goal is to collect 3 treasures.
Surprisingly, examining the database file for the game reveals
that it includes most of the scripted actions present in
the full game and placeholders for many of the rooms and items,
with their descriptions replaced with ``.''.

\section{Adapting the Game Engine}
\label{sec:adapt}

We sought to find a solution to a SAGA game using the bounded model-checker CBMC.
We chose CBMC for two reasons.
Firstly, it is a relatively mature tool with reasonable documentation;
unusually for a verification tool, it is now included in the stable release of Debian Linux.
Secondly, bounded model-checkers tend to be very good at finding counter-examples where a significant amount of low-level reasoning is required,
which we judged would be important in this case.
We also tested CPAchecker~\cite{DBLP:conf/cav/BeyerK11}, which came second in the ReachSafety category in SV-COMP 2020.
CPAchecker can be configured to use a range of techniques, including bounded model-checking and predicate analysis.
The tool that came first, VeriAbs, uses a portfolio of other tools, including CBMC, so we did not think it would add any further insights.

We now explain in detail how we adapted the ScottFree interpreter,
firstly to pose solution of a game as a C program verification problem,
and secondly to transform that problem into one solvable by CBMC.
The interpreter consists primarily of a single C file, which is 1454 lines long
(or 1358 SLOC according to \texttt{sloccount}).
Table~\ref{tab:functions} shows the size and purpose of different functions
in the file.

\begin{table}[b!]
\caption{Functions in the ScottFree interpreter.}
\label{tab:functions}
\begin{tabular}{|p{11em}|r|p{23em}|}
\hline
\emph{Functions} &
\emph{SLOC} &
\emph{Purpose} \\
\hhline{|=|=|=|}
\raggedright
strncasecmp
Aborted
Fatal
RandomPercent
ClearScreen
MemAlloc
&
56
&
Miscellaneous utility functions.\\
\hline
\raggedright
CountCarried
&
12
&
Iterates through all items in the game and counts those carried by the player.\\
\hline
\raggedright
MapSynonym
MatchItem
&
33
&
Given a noun and the player's location, returns the ID of the item referred to by the noun.\\
\hline
\raggedright
ReadString
LoadDatabase
&
156
&
Loads a game database from a data file.\\
\hline
\raggedright
OutReset
Look
OutBuf
Output
OutputNumber
&
165
&
Describes the player's location and displays messages about in-game events.\\
\hline
\raggedright
WhichWord
LineInput
GetInput
&
116
&
Reads 2-word verb/noun commands from standard input and matches them against a list in the game database.\\
\hline
\raggedright
SaveGame
LoadGame
&
52
&
Loads or saves mutable parts of the game database.\\
\hline
\raggedright
PerformLine
&
368
&
Checks the guard of a scripted event (up to 5 conditions). If it holds, executes the event (up to 4 actions).\\
\hline
PerformActions
&
200
&
Iterates through all scripted events and either executes all automatic events,
or executes those matching the verb/noun trigger entered by the player.
Also handles player movement and taking/dropping items (actions common to all games).\\
\hline
main
&
150
&
Repeatedly reads and processes a line of player input.
Also reduces duration of temporary light sources.\\
\hline
\end{tabular}
\end{table}

We aimed to modify the interpreter as little as possible for two reasons.
Firstly, we wanted to ensure that the problem solved by the tool was as accurate a representation of the original problem as possible:
human intervention introduces the possibility of error.
Secondly, we wanted to showcase the power of fully automated verification tools:
human intervention demonstrates a weakness in the tools.
As a result, we often found that our initial attempts to adapt the interpreter were inadequate,
which led us to make further adaptations later on.
As we focused on the use of CBMC, which relies on unrolling loops, we most often became aware that this was necessary by watching it log its attempts to unroll loops.
One common problem was that CBMC was unable to infer a bound for the number of iterations of a loop and unrolled it repeatedly with no sign of stopping.
Another was that there were several large, nested loops, which slowed down CBMC's translation intolerably,
and would most likely result in a giant SAT instance that could not be solved in a reasonable amount of time.

\paragraph{Asserting Unsolvability}

The first step was to insert an assertion stating that a game could not be won.
In some games, the objective
is to find a number of pieces of treasure and move them to a fixed ``treasure room''.
The player wins the game by running the ``SCORE'' command in this location,
which causes the game to print a congratulatory message and terminate.
In games without treasure, victory and defeat are both indicated
by the game ending, which is triggered by a scripted event.

To support both styles of game,
we asserted that the code to end the game via a scripted event,
located in \texttt{PerformLine}, was unreachable.
We generated a distinct assertion for each line of scripted code that ended the game.
The distinction between winning and losing is usually obvious to a human
player from the messages displayed,
but it is otherwise not indicated at the level of the game's scripting language.
Therefore a human will have to review the plays of the game generated by the model-checker
to determine which result in victory and which result in defeat.
The number of scripted events that end the game is usually small (maybe 10),
so this is not onerous.

\paragraph{Fixing the Game}

The function \texttt{LoadDatabase} reads a game from a text file,
which encodes the game as a sequence of integers and fixed strings.
CBMC and other model-checkers usually abstract file reads as nondeterminism.
In this case, we wanted the file to be fixed to a specific game,
so we initially replaced calls to \texttt{fscanf} with calls to a function that
returned in sequence the integers and strings from the file.

This worked, but unrolling of the loop in \texttt{LoadDatabase} was very slow.
So instead, we modified a copy of the interpreter (which we called \texttt{dumper}) to dump a C header file
\texttt{db.h} immediately
after reading the game; when included by the original interpreter,
\texttt{db.h} initialised the arrays that held the game database with the
relevant constants. This allowed us to remove \texttt{LoadDatabase} entirely.

As the game engine was not being used interactively,
we also had no need for \texttt{SaveGame} and \texttt{LoadGame}, so we removed those too.

\paragraph{Removing Display Code}

CBMC was unable to infer loop bounds for much of the code in \texttt{Look} that displayed the description of the player's current location.
The code only modifies local variables and makes calls to the terminal
library to move the on-screen cursor or print strings.
This clearly did not alter the game state, so we removed it.
For the same reason, we removed calls to \texttt{Output} and \texttt{OutputNumber},
whose only purpose was to print strings.

\paragraph{Bounding Player Input}

The interpreter has a \texttt{while} loop in \texttt{main} that repeatedly describes the player's current location,
asks for input and interprets it.
We replaced this with a bounded \texttt{for} loop,
limiting CBMC's search to solutions to the game below a fixed number of moves.

Commands from the player are read into a buffer before being parsed by the interpreter
in the functions \texttt{GetInput} and \texttt{LineInput}.
The buffer is large (256 bytes) but fixed-size, so each command introduced 2,048 nondeterministic bits.
However, SAGA games only permit commands of one or two words (a verb and noun) and only consider the first 3 characters of each.
Thus we could safely replace the input routine with a non-deterministic assignment to the first 7 characters and a terminating zero,
reducing this to 56 bits, or 35 bits if we allowed only upper-case characters and space.

This is still quite large and CBMC was unable to infer a bound for unrolling the use of \texttt{scanf} to lex the commands.
So instead we removed \texttt{GetInput} entirely and nondeterministically set the verb and noun to an integer index from the game's fixed list of verbs and nouns,
reducing the choice of verb and noun to 16 bits of nondeterminism.
We used CBMC's \texttt{\_\_CPROVER\_INPUT} function to log the words chosen in any counterexample trace.

There was a slight complication in that the code for taking an object in \texttt{PerformActions} passes the name of the noun for the item, not its index,
to \texttt{MatchUpItem} in order to check whether it is in the current room and available to pick up.
We initially resolved this using an \texttt{assume} statement to enforce that the name of the item picked up was equal to the noun.
However, on reflection, we realised this was unsound: normally, if there were two items with the same name in the same room,
the player would only be able to pick up the one with the numerically lowest ID, but our modification allowed either to be picked up.
So instead we adapted \texttt{dumper} to generate an equivalent function in
\texttt{db.h} that took a noun index instead of a noun string.
Fig.~\ref{fig:matchup} shows an example function.

\begin{figure}[t]
\scriptsize
\begin{verbatim}
// Original function:
int MatchUpItem(char *text, int loc)
{
    char *word=MapSynonym(text);
    int ct=0;
        
    if(word==NULL)
        word=text;
        
    while(ct<=GameHeader.NumItems)
    {
        if(Items[ct].AutoGet && Items[ct].Location==loc &&
            strncasecmp(Items[ct].AutoGet,word,GameHeader.WordLength)==0)
            return(ct);
            ct++;
    }
    return(-1);
}

// Specialised implementation for Tutorial 4:
int MatchUpItem(int no, int loc) {
    switch (no) {
        case 7: // KEY
            if (ItemLocs[3] == loc) return 3;
                return -1;
        case 8: // COIN
            if (ItemLocs[5] == loc) return 5;
                return -1;
        case 9: // CROSS
            if (ItemLocs[1] == loc) return 1;
                return -1;
        default:
            return -1;
    }
}
\end{verbatim}
\caption{The evolution of \texttt{MatchUpItem}.}
\label{fig:matchup}
\end{figure}

\paragraph{Bounding Loops with Constants}

The interpreter includes a number of loops bounded by constants read from the database.
For example, the game includes a list of scripted events and the conditions under which they may occur, which is fixed for a particular game.
CBMC was unable to determine that the bound was constant,
so we modified \texttt{dumper} to augment \texttt{db.h} with \texttt{\#define}d constants for this and other similar values.
We replaced references to the variables holding the constant with the defined constant.
For example, the struct member \texttt{GameHeader.NumItems} becomes
the constant \texttt{MAX\_ITEM}.

\paragraph{Tracking Extra State}

By this point, CBMC was able to find bounds for all loops in the program and could attempt to unwind it.
However, the unrolling was very slow.
Two possible culprits were loops that iterated through every object in the game in order to check
how many objects the player was carrying (\texttt{CountCarried}) and how many treasures the player had found
(in \texttt{PerformLine}).
In order to mitigate this, we introduced integer variables to track these values and update them whenever an object was moved.
This increased the state space, but made unrolling faster.

Pedantically, one would wish to use CBMC to verify that the variables tracked these counts correctly,
as the interpreter supports several actions that move items in quite subtle ways.
We later discarded this adaptation in favour of unrolling the loops before passing the program to CBMC,
as described in the following section;
we did not notice a difference in performance, and preferred the smaller adaptation.

\paragraph{Removal of Recursion}

Later games in the SAGA series added a shortcut command to take all objects in the player's current location
with the command ``GET ALL'' (and a corresponding shortcut to drop everything carried with ``DROP ALL'').
This feature was retrospectively hacked into the original game engine by calling \texttt{PerformActions}
recursively to try taking every object in the player's current location and setting a flag to stop a second recursive call.
(It would have been insufficient simply to move the items into the player's inventory,
as taking some of the items may have triggered scripted actions, or may have exceeded the player's carrying capacity.)

As this significantly slowed down unrolling of \texttt{PerformActions} and was not faithful to the game engine
originally shipped with most of the games, we felt it was justifiable to remove the code handling these commands.

\paragraph{Random Number Generation}

Some automatic scripted actions in games occur randomly. Instead of being
triggered on every turn, they are triggered with a fixed percentage chance.
The function \texttt{RandomPercent} determines whether they are triggered,
using the system C library's \texttt{rand} function.
We could easily have modelled this with a nondeterministic choice.
However, we instead chose to replace it with a simple pseudorandom number generator with fixed initial seed.
This meant that, after replacing the same generator in the unmodified game engine,
we could reliably test counterexample traces generated by CBMC in a real game.

% EXAMPLE?

\paragraph{Initial Infeasibility}

Having made all these changes, we applied CBMC to our modified interpreter and the smallest released SAGA game, the Adventure Sampler.
It took about an hour to unwind just a single loop of the interpreter, which made it infeasible even to construct a solution that would find one treasure,
which takes about 10 unwindings and an invocation of a SAT solver on the (presumably gigantic) unwound program.
However, we were able to verify, by adding another assertion, that it was possible to leave the starting room of the game.
To ensure that our problems were not specific to our choice of tool,
we also tried using CPAchecker.
It was able to verify that the player could leave the starting room more quickly, but its SMT solver was unable to handle anything larger in under an hour.

At this point, we wondered if we could construct a smaller game that would be feasible.
We discovered that Mike Taylor's Scott Adams Compiler, a tool for creating SAGA games, included a number of very small tutorial games.
Tutorial 4 was the smallest game with any puzzles, but that was still infeasible.
It seemed that we needed a more drastic modification of the game engine.

% needs more factual detail

\section{Partial Evaluation of Interpreter Loop}
\label{sec:pe}

By this stage, we had removed a significant proportion of ScottFree's code. What remained was:
\begin{itemize}
\item the top-level loop from \texttt{main}, which nondeterministically ``read'' commands from the player;
\item \texttt{PerformActions} to execute built-in ``GO'', ``GET'' and ``DROP'' commands;
\item \texttt{PerformLine} to execute scripted actions;
\item \texttt{CountCarried} to check the player did not exceed his inventory capacity.
\end{itemize}

\begin{figure}[p]
\scriptsize
\begin{verbatim}

int PerformLine(int ct) {
    int param[5],pptr=0;
    int act[4];
    int cc=0;
    while(cc<5) {
        int cv = Actions[ct].Condition[cc] % 20;
        int dv = Actions[ct].Condition[cc] / 20;
        switch(cv) {
            case 0:
                param[pptr++]=dv;
                break;
            case 1:
                if(Items[dv].Location!=CARRIED)
                    return 0;
                    break;
            // cases 2-19 omitted
        }
        cc++;
    }

    act[0]=Actions[ct].Action[0] / 150;
    act[1]=Actions[ct].Action[0] % 150;
    act[2]=Actions[ct].Action[1] / 150;
    act[3]=Actions[ct].Action[1] % 150;
    cc=0;
    pptr=0;
    while(cc<4) {
        if(act[cc]>=1 && act[cc]<52) {
            Output(Messages[act[cc]]);
        }
        else switch(act[cc]) {
            case 53:
                Redraw=1;
                Items[param[pptr++]].Location=MyLoc;
                break;
            // cases 52, 54-89 omitted
        }
        cc++;
    }
    return 1;
}

int PerformActions(int vb, int no) {
    // hard-coded GO omitted
    while(ct<=GameHeader.NumActions) {
        int vv = Actions[ct].Vocab / 150;
        int nv = Actions[ct].Vocab % 150;
        if (vv==vb && nv==no) {
            if (PerformLine(ct)>0 && vb!=0)
                return;
        }
        ct++;
    }
    // hard-coded GET and DROP omitted
}

int main() {
    PerformActions(0,0);
    for (int n = 0; n < MAX_MOVES; n++) {
        int vb = nondet_uchar();
        int no = nondet_uchar();
        __CPROVER_assume(vb <= MAX_WORD);
        __CPROVER_assume(no <= MAX_WORD);
        __CPROVER_input("verb", Verbs[vb]);
        __CPROVER_input("noun", Nouns[no]);
        PerformActions(vb, no);
        // light source handling code omitted
        PerformActions(0,0);
    }
}

\end{verbatim}
\caption{The structure of the interpreter loop, simplified.}
\label{fig:loop}
\end{figure}

The core of the engine as it now stood, with some simplifications made and details omitted, is shown in Fig.~\ref{fig:loop}.
CBMC was now able to infer bounds for all loops, but was still infeasibly slow.
The main difficulty was handling scripted events that were specific to a game.
In the pseudocode of Fig.~\ref{fig:engine}, these are illustrated alongside the code for the engine,
as if they were written in the same language as part of the same program.
In reality, they are encoded using a simple scripting language as part of the game database,
and a large part of the game engine is an interpreter for this language.

Although the language is structurally very primitive,
effectively limiting script programs to a sequence of \texttt{if} statements,
it features a large number of conditions that can be checked (around 20 distinct conditions, some with parameters)
and actions that can be performed (around 40, again with parameters).
The interpreter for one line of script, \texttt{PerformLine}, consists of two large \texttt{switch} statements
(one for checking conditions and one for performing actions),
each in a loop
(as there may be up to 5 conditions and 4 actions).
\texttt{PerformLine} is itself called from within a loop in \texttt{PerformActions},
which iterates through and executes each line of the script program in sequence.

For a bounded model-checker such as CBMC, the problem here is that,
while the number of dynamic paths through the script program may be relatively constrained,
it is difficult to determine this statically from the source code for the interpreter.
To do so requires knowing that:
\begin{itemize}
\item the script program, stored in the array \texttt{Actions}, is constant;
\item the loop counter \texttt{cc} increments on each iteration of each loop;
\item hence the values of \texttt{cv}, \texttt{dv} and \texttt{act} are determined entirely by the line \texttt{ct} and the loop iteration \texttt{cc};
\item thus the case taken in each \texttt{switch} is determined entirely by \texttt{ct} and \texttt{cc}.
\end{itemize}
It is within the ability of current static analysis technology to determine this,
but apparently not within the current capabilities of CBMC,
which ends up unrolling the loops to consider all possible branches of the \texttt{switch} statements.

The difficulty of handling an interpreter loop is not specific to CBMC or more generally to bounded model-checking.
Other program verification approaches have their own problems.
Consider, for example, the use of abstract interpretation with predicate abstraction.
Because the same block of game engine code is used to interpret every line of script program,
a tool using this approach would effectively have to encode the behaviour of the script program in its abstract domain.
This may be possible in theory, but would require infeasibly large predicates.

We mentioned that the necessary reasoning was within the ability of current static analysis.
Our solution to this problem was to use another tool to perform the necessary analysis and simplify the program before passing it to the verification tool.
We initially planned to use the partial evaluation tool LLPE~\cite{DBLP:phd/ethos/Smowton15}.
Small experiments showed that it was capable of unrolling loops, evaluating constant expressions, and removing redundant \texttt{switch} statements.
Unfortunately, it crashed when we tried to use it on the game engine.

Instead, we fell back on using the optimisation phase of the LLVM/Clang compiler.
The compiler's optimisation phase is quite capable of unrolling loops, evaluating constants,
inlining functions and eliminating redundant control flow branches, which was all we needed here.
However, even using \texttt{\#pragma unroll} and \texttt{attribute\_\_((always\_inline))},
after examining the disassembled intermediate code,
we found the compiler did not perform the required simplification reliably.
In order to achieve this, we were forced to unroll the interpreter loop semi-manually,
using some C preprocessor macros.
We once again modified \texttt{dumper} to add these macros to \texttt{db.h}
so that the unrolling process could still be automated for any game.

Now, compiling our modified game engine with manually unrolled loops,
Clang was able to identify which branch was taken in each iteration and hence remove the \texttt{switch} statements from the program.
However, the compiled program, in LLVM IR, could not be fed directly back into CBMC.
To work around this we used the LLVM C Backend~\cite{llvmcbe} maintained by the Julia project to translate LLVM IR back into C,
which could then be passed to CBMC.
Effectively, we had compiled the script program into C!

To give an idea of how much difference elimination of unused branches in the interpreter loop makes,
consider the game engine compiled and specialised for Adventureland.
With compiler optimisations turned on (and hence, the majority of branches eliminated), the resulting C program
generated is 8 kSLOC. With optimisations turned off, the program is 493 kSLOC.

For Tutorial 4, CBMC was able to translate the resulting program into a SAT instance and solve it.
Having adopted Clang into our process, we also added pragmas to ensure
that \texttt{CountCarried} and the loop to count collected treasure in \texttt{PerformLine} were unrolled.

Rather than translating from LLVM IR back to C,
we could have used a tool that works with LLVM IR,
such as the program verifier SMACK~\cite{DBLP:conf/cav/RakamaricE14} or the symbolic execution tool KLEE.
However, this would have been an unsatisfying conclusion to a case study that began by looking at tools in SV-COMP,
which specifically focuses on finding the best tools for C program verification.
For the moment, there is no competition for tools targeting LLVM IR and,
as it is a consciously unstable format (so as not to burden the LLVM developers with ensuring backwards compatibility), it seems unlikely one will be created.

There were some minor difficulties involved in using the C program generated by LLVM CBE.
One was that it risked mangling the names of CBMC's \texttt{\_\_CPROVER}
hooks unless they were declared \texttt{extern}.
Another was that it translated the strings in the \texttt{Verbs} and \texttt{Nouns} arrays into arrays.
Although the run-time behaviour of the program was unchanged,
as a string in C is just a \texttt{char} array,
CBMC no longer recognised the array members as strings, so it did not print them when they were passed to \texttt{\_\_CPROVER\_INPUT}.
The solution was to split off \texttt{main} and the definition of \texttt{Words} into a small file \texttt{stub.c} that could
be concatenated onto the end of the output from LLVM CBE before passing it to CBMC.
Then, running CBMC with \texttt{--trace},
it was possible to \texttt{grep} for lines containing \texttt{INPUT} to read off the required game moves.

Now that we were able to solve Tutorial 4 (and all the other tutorial games),
we returned to Adventureland and the Adventureland Sampler.
Although now faster, both were still infeasibly slow to unroll beyond a few moves.
CBMC also gave an error, when unrolling just one move, that it needed a higher value for the \texttt{--object-bits} setting,
indicating that it was trying to track too many regions of memory.
This was easily changed, but would not help with the speed of unrolling.

Recall that the Adventureland Sampler contained much of the database for the full Adventureland game.
Using \texttt{sac}, we deleted all items, rooms and messages with the placeholder text ``\texttt{.}'',
and all scripted actions that referred to these.
This cleaned-up version of Adventureland Sampler was small enough for our tool to handle, although it took several hours
to generate the 23-move solution.

\section{Results}
\label{sec:res}

\begin{table}[b]
\caption{Statistics for solvable games. \texttt{sampler2} is our modified version of the Adventureland Sampler.
\texttt{adv01} is Adventureland, which has the same size statistics as the unedited Sampler.}
\label{tab:solvestats}
\begin{tabular}{r|rrr|r|rrr}
Game & Items & Actions & Rooms & Moves &
\parbox{6em}{\raggedleft CBMC CPU time} &
\parbox{6em}{\raggedleft Of which solver time} &
\parbox{7em}{\raggedleft Max Res Memory (MB)} \\
\hline
t3 &	5 &	5 &	4 & 9 & 3s & 1s & 152 \\
t4 &	7 &	7 & 	6 & 14 & 25s & 5s & 412 \\
t5 &	10 &	10 &	7 & 14 & 93s & 14s & 667 \\
sampler2 & 32 & 100 & 18 & 23 & 5h 32m 34s & 1h 15m 11s & 15421 \\
adv01 & 66 & 170 & 34 & - & - & - & - \\
\end{tabular}
\end{table}

Table~\ref{tab:solvestats} shows information about the solvable games,
namely the tutorial games and our modified version of the Adventureland Sampler.
Tutorials 1 and 2 are not shown, as they cannot be won or lost.
Timings are for an Intel Core i7-7500U at 2.70 GHz with 16 GB of RAM.
Solver time is the time taken by CBMC's SAT solver.
Notice that this is only a small fraction of the total time taken by CBMC.
Timings do not include the time taken to build the specialised version of the game engine.

Solutions generated by our tool, ScAmPI, can be reproduced using the
artifact associated with this paper.
The solution generated for Tutorial 5 by a human would normally be longer.
This game includes dark rooms, where nothing is visible unless a lamp is carried.
However, the solver is quite capable of navigating through the game safely
without being able to see the rooms, so it does not pick up the lamp.

% Why does Sampler need 23 moves, not 22? Is the pause to avoid random death
% from chiggers?

\begin{figure}[p]
\resizebox{1.0\textwidth}{!}{\includegraphics{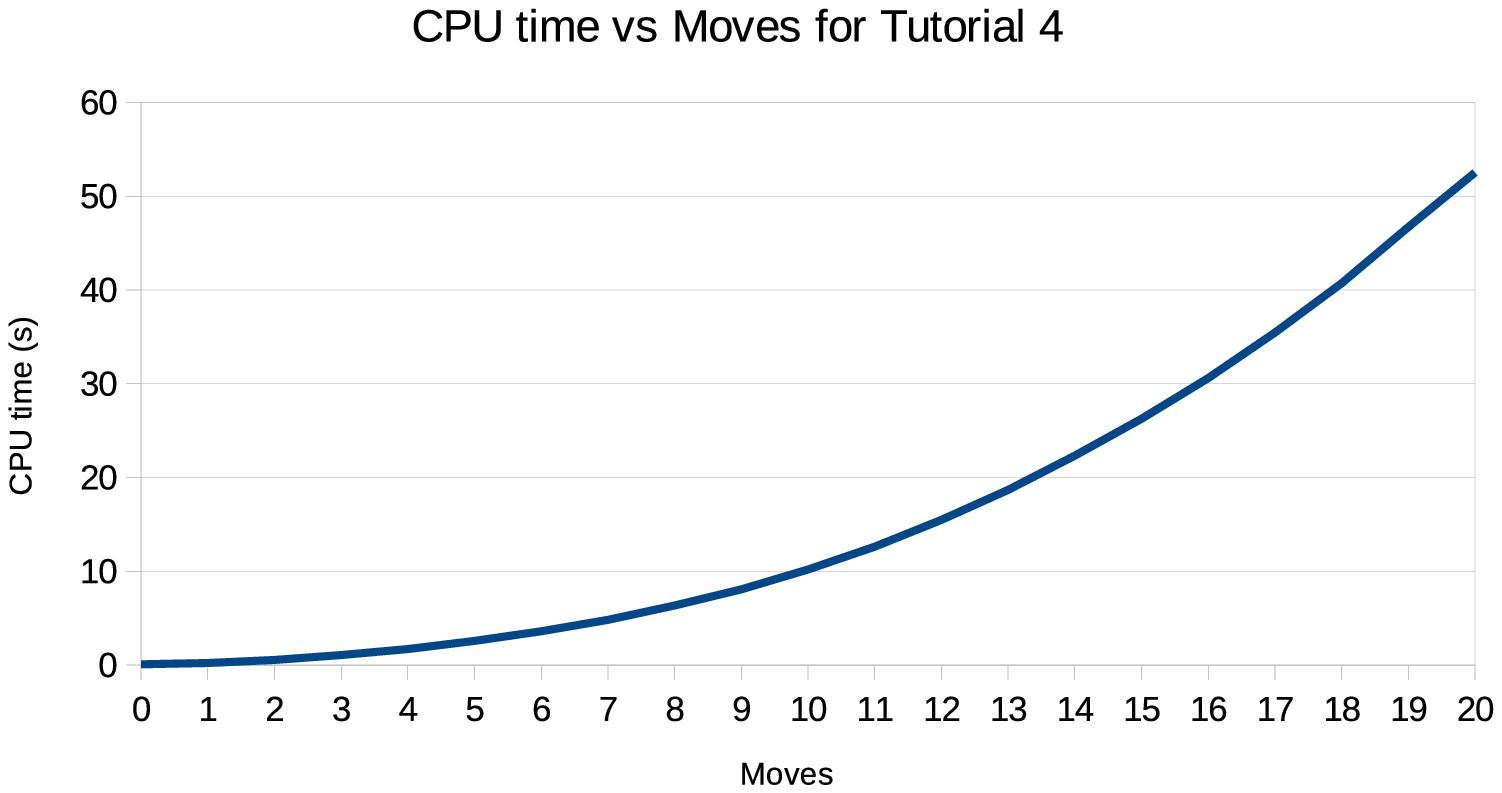}}
\resizebox{1.0\textwidth}{!}{\includegraphics{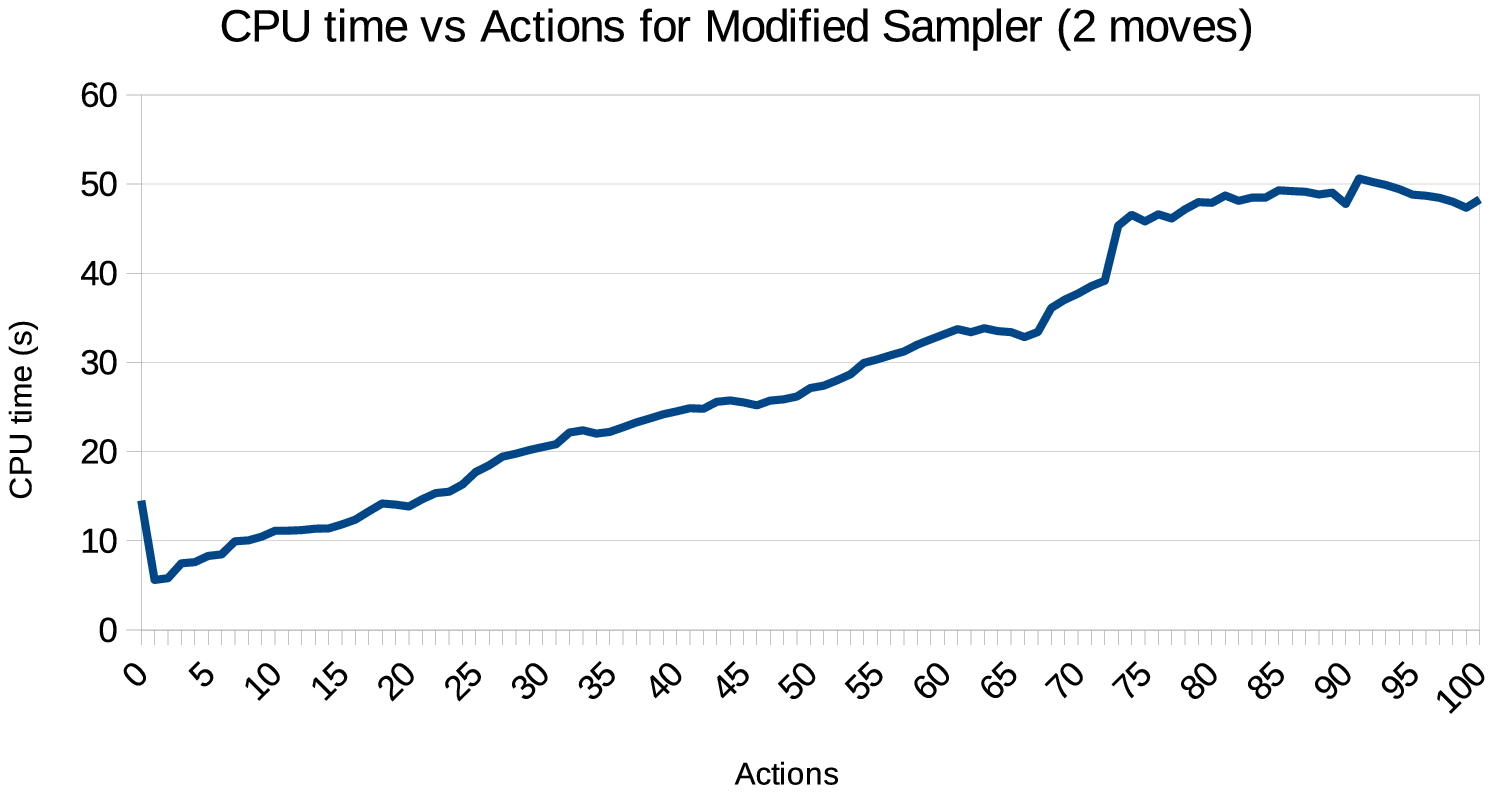}}
\caption{How time taken by CBMC (excluding time for the SAT solver) scales
with increasing numbers of moves and actions.}
\label{fig:graphs}
\end{figure}

The top graph in Fig.~\ref{fig:graphs} shows how the time taken to generate a SAT instance for the program for Tutorial 4 increases with the number of moves.
It appears to be roughly quadratic.
The bottom graph shows how the time taken for the modified Sampler changes as scripted actions are added to the game.
It is clear that adding more actions makes CBMC slower. The rate of growth seems variable, but roughly linear.

Solving the Sampler used nearly all of our test machine's memory and took several hours.
Adventureland is larger than the modified Sampler in every way. It has more items, rooms and scripted actions.
Solving the game will take more moves.
The size of the state space of the game is $\mathcal{O}(r^i)$ for $r$ rooms and $i$ items.
We have seen that the time needed grows with both the number of moves considered and the number of actions.
The other SAGA games are comparable in size to Adventureland.
Taking all this into consideration, it looks unlikely that we will be able to solve any of these games
without further optimisations or technical advances.

This may yet be possible, but would require a more detailed analysis to discover
which aspects of the game engine are contributing most to CBMC's slowdown.
Recall our earlier remark that a tool that can solve interactive fiction games
must be either very good at low-level reasoning, or able to mix both high-level and low-level reasoning.
CBMC is very good at low-level reasoning.
However, with the Sampler featuring 67 possible different verbs, and almost as many nouns,
every 2-word input command has around 12 bits of nondeterminism.
Thus to solve the game in 23 steps requires correctly resolving 276 nondeterministic bits.
The other SAGA games could require over 1,000 bits, which seems large;
it may be more productive to look at a tool with better high-level reasoning.

\section{Evaluation}
\label{sec:eval}

\paragraph{Applying our technique to other interpreted languages.}

We are optimistic that our technique can be applied to other interpreted languages.
Our work shows that it can be effective in reducing the number of paths considered
in a scripting language with a moderately large number of built-in operators or functions.
On that basis, it could also be applied to an emulator for a microcontroller.
There is a port of ScottFree to the Z80 microprocessor,
which could make for an amusing but challenging follow-up case study,
as the scripting language interpreter loop would be nested inside the emulator loop.

However, the scripting language we considered has very limited flow control.
It remains to be seen whether it is effective on languages with,
for example, while loops or computed jumps.

\paragraph{On adapting the game engine and partial evaluation of the interpreter loop.}

With the exception of replacing player input with nondeterminism,
the transformations we applied to make the game amenable to solution
with a model-checker
could mostly be summarised as follows:

\begin{enumerate}
\item Remove code that is redundant because it only displays output.
\item Where loop bounds and other values can be calculated statically, replace them with a constant.
\item Unroll loops with constant bounds.
\item If calculation of a function is costly, but could be done incrementally, add and maintain a variable that caches its value.
\end{enumerate}

These are all variants of classical compiler optimisations,
namely dead code elimination, constant folding/propagation, loop unrolling and strength reduction.
In principle, the majority of the changes we made could be automated.
Indeed, we were ultimately able to perform much of the necessary transformation using Clang,
\emph{but} some tweaks and pragma annotations were required.
If we had been able to use dedicated partial evaluation tool, like LLPE, but more mature,
the human effort required would no doubt have been smaller.

In practice, the difficulty is often in identifying where and when an optimisation or transformation is needed.
In order for our approach to be applied to other interpreters without human intervention,
we would need a way of identifying the location and structure of an interpreter loop.
In the case of a game engine, it may be possible to do this heuristically by using a form of run-time program analysis
on plays of the game with random input.

\paragraph{General reflections.}

Based on our experiences in this case study, we would make the following general recommendations
to someone interested in using C program verification tools:
\begin{itemize}
\item \emph{Use the SV-COMP results to find out which tools are strongest at the moment.} Try to identify the category that best fits your application domain.
\item \emph{Learn to interpret the logs of the tools you use.} With experience, you can make good intuitive guesses about why a tool is not working well.
\item \emph{Find the smallest representative example problems first.} If they do not exist, make them yourself.
If you have a large example that does not work, it is often hard to understand why. It is easier to make a small example larger or more complex until it breaks.
\item \emph{Benchmark modifications you make, so you can tell if they actually work.} Remember graphs are helpful in spotting trends.
\item \emph{Implement the harder/better improvement if the estimated difference in effort is small, or (initially) the easier/worse improvement if the difference is large.}
This tends to minimise wasted development effort.
\item \emph{Favour tools with good documentation.}
Although many publicly available tools are both powerful and highly configurable,
documentation, where it exists, is often very limited,
which limits their applicability in practice.
\end{itemize}
These are mostly common sense, but common sense is not always so common.

\section{Related Work}
\label{sec:related}

\paragraph{Interactive fiction games.}

We recently developed a tool called ScAmPER~\cite{DBLP:conf/tap/Lester20} that uses a heuristic-guided, explicit-state search
to analyse SAGA games.
The tool outputs a suite of game inputs that serve as test cases
to maximise coverage of lines of scripted code within a game.
It is able to achieve a high level of coverage fairly quickly,
but is unable to prove when a line of code is unreachable.
In contrast, the tool we discuss in this paper, ScAmPI, is slow and only works on smaller games,
but the C program it uses could be used (with a suitable verification tool, not a bounded model-checker)
to prove a line of code is unreachable.

ScAmPER's heuristics encode domain-specific knowledge about how to play interactive fiction games.
ScAmPI is more interesting technically because it works using program model-checkers
that have no knowledge of the application domain.
Our preliminary work on ScAmPI, showing how to solve Tutorial 4 using
CBMC~\cite{wpte},
inspired another researcher to re-implement our idea using symbolic computation on an interpreter written in Haskell~\cite{ifl}.
Neither ScAmPER nor ScAmPI could solve the original Adventureland Sampler.
ScAmPER was able to explore the majority of the modified Sampler, triggering
almost every scripted event, but was not able to complete the game,
presumably because it was unable to work out how to get all 3 treasures simultaneously.
As ScAmPI was able to solve this game, it constitutes both a technical improvement
and, in at least this one case, a practical improvement.

See our previous work~\cite{DBLP:conf/tap/Lester20} for a summary of other research on automated solution of interactive fiction games.
Of greatest significance is Pickett, Verbrugge and Martineau's work on model-checking SAGA games~\cite{pickett2005nfg} using NuSMV.
Their approach works by translating the game into a high-level formalism called PNFG (Programmable Narrative Flow Graph),
which is compiled to a lower-level formalism based on Petri nets.
However, there is no indication of whether or how the translation to PNFG was automated,
which raises concerns that a human translator may have simplified the problem,
or that the translation, even if automated, may have introduced an error.
In contrast, our approach, which uses the game engine's source code,
gives a high degree of confidence that the problem solved by the model-checker is an accurate representation of the original game.

\paragraph{Model-checking other games.}

Model-checking techniques have been applied to a range of other computer puzzle games.
Kwon~\cite{DBLP:conf/atva/KwonL04} considers encoding and solving Sokoban puzzles using NuSMV.
Like interactive fiction games, Sokoban involves moving a player and discrete objects around discrete locations on a map,
so it also suffers from state space explosion.
Moreno-Ger and others~\cite{DBLP:journals/infsof/Moreno-GerFSF09}
consider encoding point-and-click adventures written for the engine eAdventure
and using NuSMV to check that they have no dead ends.
These games do not suffer from state space explosion to the same extent,
as objects cannot usually be dropped after they are picked up,
so they are either in their initial location, held by the player, or no longer in the game.
Like interactive fiction games, they usually feature a limited form of scripting.
In both of these cases, the translation from the puzzle or game instance to NuSMV is automated,
but the translation was designed manually.

Model-checking has been used to find bugs in commercial games.
Hasegawa and Yokogawa~\cite{DBLP:conf/icfem/HasegawaY19} use NuSMV to find likely bugs in scripted
events in the game Final Fantasy XV.
However, to our knowledge, there is no realistic previous work that takes the code of a game engine as its starting point.
Teeny Tiny Mansion~\cite{tttm} is a toy example of an interactive fiction game whose
source code can be verified to have no dead ends using CBMC,
but it was written specifically for this purpose.

\paragraph{Partial evaluation and interpreted languages.}

Partial evaluation of programs has been studied extensively as a method of optimisation.
The idea of using a partial evaluator on the combination of an interpreter and a program in an interpreted language
is often attributed to Futamura~\cite{DBLP:journals/lisp/Futamura99a}.
The literature on partial evaluation suggests this as one way of producing an optimising compiler~\cite{DBLP:books/daglib/0072559}.
Jones suggests how best to write an interpreter so that it is amenable to partial evaluation~\cite{DBLP:journals/scp/Jones04}.

Partial evaluation has been suggested as a way of improving the effectiveness of verification techniques~\cite{DBLP:journals/csur/DwyerHN98},
but does not seem to have gained much prominence,
and much of the work has been in the context of functional or logic programming, which makes it difficult to apply to interpreters written in C.
For example, Lisitsa and Nemytykh~\cite{DBLP:journals/corr/abs-1708-09002} verify cache coherence protocols
in an interpreted Lisp-like language.
Meanwhile, Amin and Rompf~\cite{DBLP:journals/pacmpl/AminR18} establish the necessary theory for removing multiple nested levels of interpretation
in the context of staged lambda calculus.

The problem of applying verification tools to new and exotic programming languages is well-recognised.
A common approach, for example used by SMACK~\cite{DBLP:conf/cav/RakamaricE14} and LLBMC~\cite{DBLP:conf/vstte/MerzFS12},
is to target a low-level intermediate language, such as LLVM IR.
Assuming the new language compiles to LLVM IR, this makes it easy to write a new verification tool and
to ensure that it faithfully models the run-time behaviour of the new language.
This approach can also be extended to verification problems involving programs written in multiple programming languages~\cite{DBLP:conf/vmcai/GarzellaBHR20}.
The problem has not been studied so well for interpreted languages.
Bucur and others~\cite{DBLP:conf/asplos/BucurKC14} develop a framework for creating a symbolic execution engine for an interpreted language, using the interpreter.
Like us, they observe the problem of distinct lines of high-level code using the same lines of low-level code in the interpreter loop.
They identify strategies for minimising the impact of this and recovering the high-level structure of the interpreted program
through instrumentation of the interpreter.
Porncharoenwase and others~\cite{DBLP:conf/vmcai/Porncharoenwase20} propose an approach called SymFix for tackling path explosion in symbolic execution,
which they demonstrate on an interpreter for a subset of the ARM instruction set.

\section{Conclusions}
\label{sec:conc}

Solving old computer and video games is an appealing application of formal verification
with relevance to current problems.
By applying program transformations and partial evaluation to a game engine,
we were able to solve a small, commercially released interactive fiction game by using
the C program model-checker CBMC on the game engine.
We are unsure whether our technique can scale further, to solving larger games;
we suspect the exponential growth of the state space with number of items will make this difficult, but we invite other researchers to try.
We hope that our case study and insights will be instructive to other users of verification tools
and motivate future developments in verification tools.

Looking beyond this case study,
it would be useful if more of the transformation steps could be fully automated
within an existing verification tool or framework.
The main technical difficulty here would likely be
how to identify the structure of an interpreter loop, so that the transformations could be applied
in the right place.

\paragraph{Acknowledgements.}
We are grateful to Christopher Smowton for his comments about LLPE, LLVM CBE and CBMC.

\bibliographystyle{splncs04}
\bibliography{refs}
\end{document}